\documentclass[preprint,showpacs,preprintnumbers,amsmath,amssymb]{revtex4}

\usepackage{graphicx}% Include figure files
\usepackage{dcolumn}% Align table columns on decimal point
\usepackage{bm}% bold math

\begin{document}

%\preprint{subbmitted to Phys. Rev. Lett. (2005)}

\title{Resonant pairing between Fermions with unequal masses}
% Force line breaks with \\

\author{Shin-Tza Wu}
\author{C.-H. Pao}
\affiliation{%
Department of Physics, National Chung Cheng University, Chiayi
621, Taiwan
}%

\author{S.-K. Yip}
\affiliation{ Institute of Physics, Academia
Sinica, Nankang, Taipei 115, Taiwan}%

\date{\today}% It is always \today, today,

\begin{abstract}
We study via mean-field theory
 the pairing between Fermions of different masses,
especially at the unitary limit.  
At equal populations, the thermodynamic properties
are identical with the equal mass case provided an
appropriate rescaling is made.  At unequal populations,
for sufficiently light majority species, the system
does not phase separate.  For sufficiently heavy
majority species, the phase separated normal phase
 have a density larger than that of the superfluid.
For atoms in harmonic traps, the density
profiles for unequal mass Fermions can be 
drastically different from their equal-mass counterparts.
\end{abstract}

\pacs{03.75.Ss, 05.30.Fk, 34.90.+q}
\maketitle
%\section{\label{sec:level1}First-level heading:\protect\\ The line
%break was forced \lowercase{via} \textbackslash\textbackslash}

%%%%%%%%%%%%%%%%%%%%%%%%%%%%%%%%%%%%%%%%%%%%%%%%%%%%%%%%%%%%5

Cooper pairing between Fermions due to
the presence of an attractive interaction
 is the essential physics 
for most, if not all, superconductors.
In recent years, Feshbach resonances \cite{Feshbach} 
provide us with a wonderful opportunity
to further study strongly interacting Fermi systems.  
For wide resonances, the Feshbach resonance between the atoms
can be viewed as providing an attractive pairing interaction,
with the pairing strength increases from positive detuning
towards negative detuning.  
For the case of equal populations of two Fermionic species 
and an s-wave interaction between them, it was
predicted long time ago \cite{EL}
that the system can smoothly crossover from a BCS superfluid
to a Bose Einstein condensate of tightly bound pairs.
Recent experiments
have provided a confirmation of this picture,
and many thermodynamical and dynamical properties have been measured.
\cite{expr}
There is also an explosion of theoretical activities 
studying this system. \cite{theo}

In the case where the populations of the Fermions are unequal,
the smooth crossover discussed above is known to 
be destroyed \cite{untheo}.  
Recent experiments \cite{unexpr}
have shown that phase separation occurs near resonance,
so that the superfluid appears near the center of
the trap and a normal fluid of unequal population density
appears outside the superfluid core. \cite{traptheo}

In the literature cited above, the Fermi atoms involved
are identical except their internal electronic states.
Hence they have identical masses.  However,
Feshbach resonances are not confined to the same atom
species 
\cite{hFesh}.
This thus raises the possibility of studying
Fermion pairing between two atomic species of
very different masses, with either equal or
unequal populations.  
This problem can also have important implications
in other areas of physics such as quark matter
\cite{Forbes05}
and superconductivity in general.

In this paper, we thus study two Fermionic species of unequal
masses with an interspecies short-ranged s-wave attractive
interaction.  First we consider the equal population
case.  We show that, at zero temperature and
within mean-field theory \cite{EL}, the thermodynamic
properties of this system have a one-to-one correspondence
with those of the equal mass case, provided we
rescale the energies appropriately.  Next we consider
the case of unequal population, but confining ourselves
to the unitary regime where the scattering length
diverges.  Defining $\gamma$ as the ratio of the 
majority atom mass to the minority mass, we show that
the system phase separates for $\gamma > $
$\gamma_H \approx 0.13$ but remains homogeneous
for $\gamma < \gamma_H$.  
For sufficiently light majority particles, it is energetically
costly to phase separate out the majority particles
since a high Fermi energy of the majority species will result.
For $\gamma > \gamma_H$,
we show that the system phase separates into a superfluid
with equal population density and a normal fluid with unequal
population density.  However, there is dramatic difference
between the cases where $\gamma$ is larger or less
than a value $\gamma_* \approx 3.9$.  For 
$\gamma_H < \gamma < \gamma_*$, the density of
the normal fluid is less than the superfluid,
as one may expect due to the attractive interaction.
However, for $\gamma > \gamma_*$, the normal fluid
is actually {\em denser} than the superfluid.
We explain this by the larger density of states
for the heavy majority component.  
We illustrate this difference by calculating
the density profile of an unequal population gas
in an harmonic trap.

We shall then consider at zero temperature two Fermionic species with
a short-ranged interaction characterized by
the s-wave scattering length $a$.
$1/a$ increases from $- \infty$ to $+\infty$ with
the strength of the pairing interaction.
We shall call the two species up and down,
with masses $m_{\sigma}$, $\sigma = \uparrow$ or 
$\downarrow$.
The effective mean-field Hamiltonian is
$H_{mf} - \mu_{\uparrow} N_{\uparrow} - \mu_{\downarrow} N_{\downarrow}
 = \sum_{\vec k, \sigma}  \xi_{\sigma}(k) c^{\dagger}_{\vec k, \sigma}
   c_{\vec k, \sigma} 
   - \sum_{\vec k} [\Delta^* c_{- \vec k, \downarrow} c_{\vec k, \uparrow}
  + c.c]
   - [\frac{m_r}{ 2 \pi \hbar^2 a} - \sum_{\vec k} 
    \frac{2 m_r}{k^2} ]|\Delta|^2 $.
Here $\xi_{\sigma}(k)\, =\, \hbar^2 k^2/2m_{\sigma} - \mu_\sigma$ are
the quasi-particle excitation energies for normal fermions
relative to their respective chemical potentials $\mu_{\sigma}$,
 $c_{\vec k, \sigma}$'s are the annihilation operators for
the Fermion $\sigma$ at momentum $\vec k$,
{\it c.c.} denotes the complex conjugate,  
$m_r$ is the reduced mass
($m_r^{-1} = m_{\uparrow}^{-1} + m_{\downarrow}^{-1}$).
$\Delta$, the off-diagonal pairing potential,
 is determined by
minimization of the expectation value of this Hamiltonian
({\it i.e.} the free energy \cite{free})
with respect to $\Delta$.
This gives, for equal populations, 
(see, e.g \cite{Wu03})

\begin{equation}
 - \frac{  m_r }{ 2 \pi \hbar^2 a} \Delta \, =\, 
\Delta
\frac{1}{V} \sum_{\vec k}
\left [ \frac{ 1 }{ E_\uparrow (k) +
E_\downarrow (k) }\, -\, \frac{ 2 m_r }{ \hbar^2 k^2} \right ]\ .
\label{eD}
\end{equation}
$E_{\sigma}(k)$, the quasiparticle energies at wavevector $k$, are given by
\begin{equation} 
E_\sigma (k)\, = \, \frac{
\xi_{\sigma}(k) - \xi_{- \sigma}(k)}{ 2}\ +\
\sqrt{ \left ( \frac{ \xi_{\sigma}(k) + \xi_{-\sigma} (k)}
{2 } \right )^2\, + \, \Delta^2}\ ,
\end{equation}
where and $-\uparrow \equiv \downarrow$.
Defining $\mu \equiv (\mu_{\uparrow} + \mu_{\downarrow})/2$
and $h \equiv (\mu_{\uparrow} - \mu_{\downarrow})/2$ and
noticing that 
$ \xi_{\sigma}(k) + \xi_{-\sigma} (k) = $
$\frac {\hbar^2 k^2} {2 m_r} - 2 \mu$,
one obtains
\begin{equation}
E_{\uparrow}(k) + E_{\downarrow}(k) = 
 2  \sqrt { (\frac{\hbar^2 k^2}{4 m_r} - \mu )^2 + |\Delta|^2 } 
\equiv 2 \tilde E(k) \ .
\label{Esum}
\end{equation}
The number equation is 
\begin{equation}
n = \frac{1}{V} \sum_k v^2(k)
\label{density}
\end{equation}
where $v(k)$, the usual coherence factor,   is
given by \cite{Wu03}
$v^2(k)= \frac{E_{\uparrow} - \xi_{\uparrow}}
             {E_{\uparrow} + E_{\downarrow}}
      = \frac{E_{\downarrow} - \xi_{\downarrow}}
             {E_{\uparrow} + E_{\downarrow}} $
and hence
\begin{equation}
v^2(k) = \frac{1}{2} \left(
   1 - \frac{\hbar^2 k^2/4 m_r - \mu}
  {  \sqrt { (\frac{\hbar^2 k^2}{4 m_r} - \mu )^2 + |\Delta|^2 }} \right)
\label{v2}
\end{equation}
The equations (\ref{eD}) and (\ref{density}) determine
$\Delta$ and $\mu$ for a given density.  
Using eqs (\ref{Esum}) and (\ref{v2}), we see that
these equations are identical to those \cite{EL} of equal masses
$m_{\uparrow} = m_{\downarrow} = m$
provided we substitute $m \to 2 m_r$.
Thus their solutions are identical after this replacement.
(The solutions are thus independent of $h$ so long as
all quasiparticle energies $E_{\sigma} (k) > 0$:  
%see eq (\ref{Du}) below.  
This condition also 
guarantees that $n_{\uparrow} = n_{\downarrow}$, see below).
Therefore their solutions can be expressed
as 
\begin{equation}
\mu = \frac{\epsilon_{Fr}}{2}
    \zeta_1 \left( \frac{1}{k_F a} \right)
\label{zeta1}
\end{equation}
\begin{equation}
\Delta = \frac{\epsilon_{Fr}}{2}
    \zeta_2 \left( \frac{1}{k_F a} \right)
\label{zeta2}
\end{equation}
where the dimensionless functions
$\zeta_{1,2}$ are identical to the corresponding
ones for equal masses \cite{EL}.
Here $\epsilon_{Fr} \equiv k_F ^2 / 2 m_r$ and
$k_F \equiv ( 3 \pi^2 n)^{1/3}$,
where $n$ is the total density.
We have introduced a factor of $1/2$ in 
eqs (\ref{zeta1}) and (\ref{zeta2}) since
$\epsilon_{Fr}/2 = k_F^2 / (2m) $ if $m_{\sigma} = m$.
It follows immediately that other thermodynamical properties
can be obtained similarly.  For example, the energy
density $E_S/V$ can be written as 
$E_S/V = n \frac{\epsilon_{Fr}}{2} \zeta_3  \left( \frac{1}{k_F a} \right)$
where $\zeta_3$ is again the same function as the equal mass case.

In particular, at unitarity ($a = \infty$), we have
$\mu = \frac{\epsilon_{Fr}}{2} \zeta_1(0)$
where $\zeta_1(0) \approx 0.59$ in the present theory
(and thus $\mu > 0$).
The magnitude of the order parameter 
$\Delta = \frac{\epsilon_{Fr}}{2} \zeta_2(0)$
with $\zeta_2(0) \approx 0.68$ and hence 
$\Delta/\mu \approx 1.16$.  Using the fact that
the only length scale at unitarity is $k_F^{-1}$,
$\zeta_3 (0) = \frac{3}{5} \zeta_1(0)$ and
thus
$E_S/V = \frac{3}{5} n \frac{\epsilon_{Fr}}{2} \zeta_1(0)$.
At chemical potential $\mu$,  from eq (\ref{zeta1}),
$k_F = $ $ \left[ 4 m_r \mu / \zeta_1(0) \right]^{1/2}$,
hence the density is given by
$n = \frac{1}{ 3 \pi^2} \left[ 4 m_r \mu / \zeta_1(0) \right]^{3/2} $.
Since the free energy density is
$\Omega_S /V = E_S/V - \mu n = - \frac{2}{5} n \mu$, therefore 
\begin{equation}
\frac{\Omega_S}{V}
= - \frac{2}{15 \pi^2}
  \frac{ ( 4 m_r )^{3/2} \mu^{5/2} }{\zeta_1^{3/2}(0) }  \ .
\label{Omega}
\end{equation}

Mean field theory is known to produce a larger $\zeta_1(0)$
than Monte-Carlo and experiments ($\approx 0.44$) for the equal mass case
\cite{expr,theo}.
It is however of interest to see to what extent 
$\zeta_1$ etc will depend on $\gamma$
in the unequal mass case in more accurate calculations.
 
\begin{figure}[tbh]
\begin{center}
\includegraphics[width=3in]{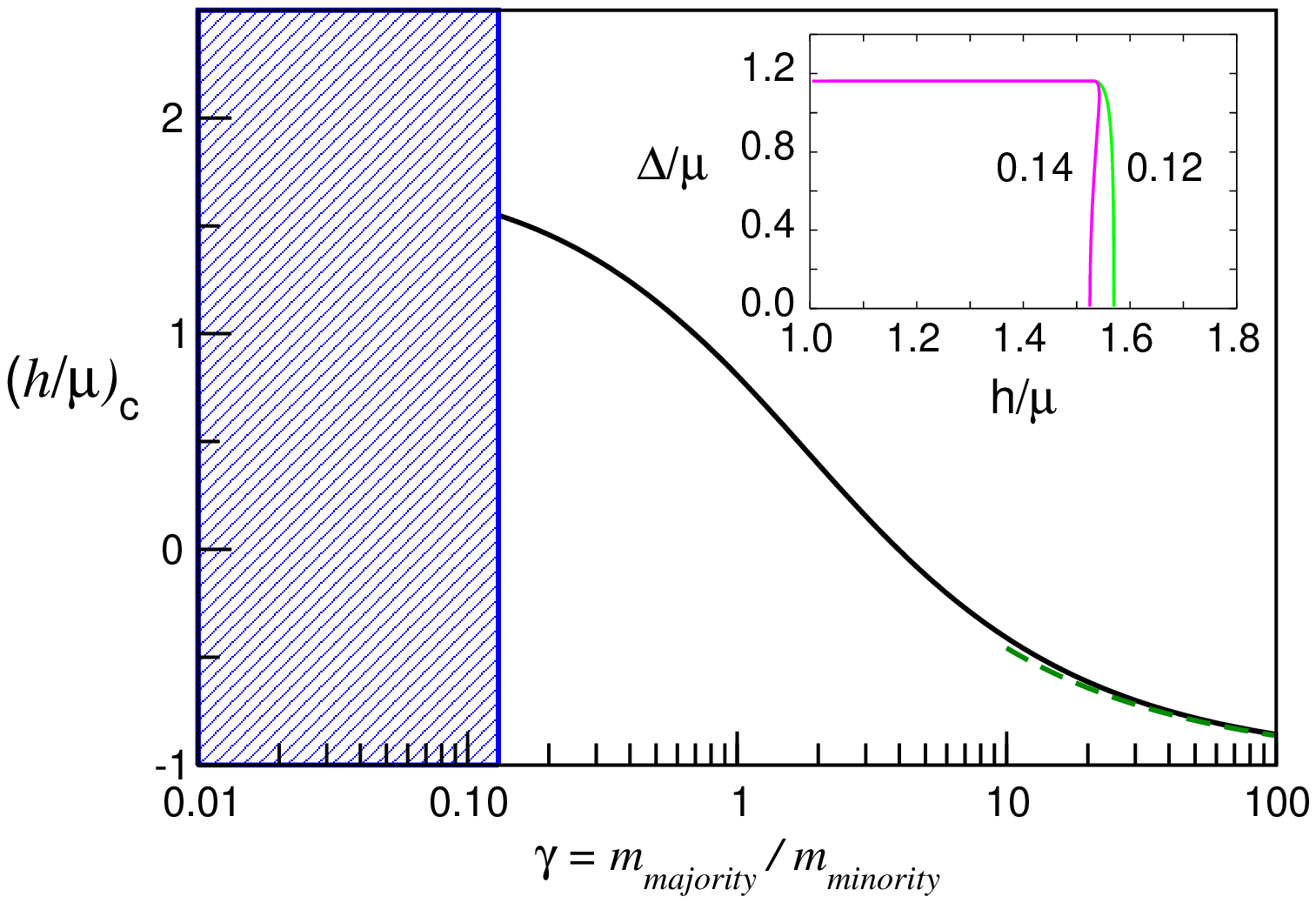}
\end{center}
 \caption{(color online) $(h/\mu)_c$ as a function of 
  $\gamma$ for $\gamma > \gamma_H$.
The dashed line is the asymptotic formula in text.
Inset:  $\Delta/\mu$ as a function of $h/\mu$ for
$\gamma = 0.14$ and $\gamma= 0.12$.}
 \label{fig:Dh}
 \vspace{-5pt}
\end{figure}

Next we consider unequal populations but confine ourselves
to unitarity ($1/a = 0$).  
We shall call $\uparrow$ ($\downarrow$) the majority (minority) species.
The self-consistent order parameter
equation reads

\begin{equation}
 0 = 
\frac{\Delta}{V}
\sum_{\vec k}
\left [ \frac{ 1 - f(E_\uparrow(k)) - f(E_\downarrow(k))}
{ E_\uparrow (k) +
E_\downarrow (k) }\, -\, \frac{ 2 m_r }{ \hbar^2 k^2} \right ]\ .
\label{Du}
\end{equation}
Due to the appearance of the Fermi functions $f$ with arguments
$E_{\sigma} (k)$, the solution to this equation does
{\em not} depend on the reduced mass alone.  
($n_{\uparrow} > n_{\downarrow}$ requires that 
some $E_{\uparrow}(k)$'s are negative,
since $n_{\uparrow} - n_{\downarrow}
= \frac{1}{V} \sum_{\vec k} f(E_{\uparrow})$,
 see, e.g. \cite{untheo,Wu03}).
By rescaling
all energies to $\mu$, it is easy to see that
eq (\ref{Du}) gives $\Delta/\mu$ as a function of $h/\mu$ for
each $\gamma \equiv m_{\uparrow}/m_{\downarrow}$.  
The non-trivial ($\Delta \ne 0$) solutions to eq (\ref{Du})
consists of two branches (see inset of Fig \ref{fig:Dh}).
One branch has $\Delta/\mu \approx 1.16$ independent of
$h/\mu$ and $\gamma$.  This branch corresponds to
$E_\sigma(k) > 0$ for all $k$ and $\sigma$, {\it i.e.},
no quasiparticles (and hence equal populations).
  This solution has already been described
in the last paragraph above.  Another branch,
corresponding to $n_d = n_{\uparrow} - n_{\downarrow} \ne 0$, gives
a $h/\mu$ and $\gamma$ dependent $\Delta/\mu$.  
For $\gamma < 0.13$, $\Delta/\mu$ decreases
with increasing $h/\mu$.  
We have verified that this branch is a free energy \cite{free}
minimum, and thus represents a stable homogeneous
superfluid. 
(We have also verified that this state has only one Fermi surface,
and hence it does {\em not} correspond to a breached-pair state
\cite{Forbes05})
 However,
for $\gamma > 0.13$, $\Delta/\mu$
increases with increasing $h/\mu$
(and correspondingly, for some range of $h$, $\Delta/\mu$
is multi-valued).  This branch
is a generalization of the solution obtained by Sarma \cite{Sarma}
in the BCS limit for $\gamma = 1$. We found that
this solution corresponds to a free energy \cite{free} relative maximum
and is thus unstable.

For a homogeneous system with $\gamma > 0.13$,
the likely scenario is that the system phase separates,
as \cite{unexpr}
in the case of $\gamma = 1$.
(We ignore the more exotic possibilities \cite{exotic} here).
To characterize this phase, we seek the point
where the free energy density of the 
completely paired superfluid state is
equal to that of the normal fluid, {\it i.e.},
$\Omega_S (\mu, h)/V  = \Omega_N (\mu,h)/V$.
$\Omega_S/V$ is independent of $h$ and was already given
before.  Within our mean field theory, $\Omega_N/V$ is just the 
free energy density of an non-interacting Fermi gas, thus
$\Omega_N/V = - \frac{1}{15 \pi^2} $
$ \left[ ( 2 m_{\uparrow})^{3/2} (\mu + h) ^{5/2}
  + ( 2 m_{\downarrow})^{3/2} (\mu - h) ^{5/2} {\cal H} 
   \right] $
where ${\cal H} = 1$ if $\mu - h > 0$ and $=0$ otherwise.
Equating the free energies, we obtain an equation for 
$h/\mu$.  Denoting this value as $(h/\mu)_c$,
we get  (using $\mu > 0$ at resonance)
\begin{equation}
\frac{1}{\zeta_1^{3/2}(0)} =
\frac{1}{2} \left[
   \left( \frac{1 + \gamma}{2} \right)^{3/2}
    \left( 1 + \left( \frac{h}{\mu} \right)_c \right)^{5/2}
  +
 \left( \frac{1 + \gamma^{-1}}{2} \right)^{3/2}
    \left( 1 - \left( \frac{h}{\mu} \right)_c \right)^{5/2} {\cal H}
  \right]
\label{eqm}
\end{equation}

For $h/\mu < (h/\mu)_c $, we have a paired superfluid, whereas
for $h/\mu > (h/\mu)_c$, we have a normal fluid.
At $h/\mu = (h/\mu)_c$, the system phase separates into a
superfluid region and a normal region.  The dependence
of $(h/\mu)_c$ on $\gamma$ is given in the main part of
Fig \ref{fig:Dh}.  
$(h/\mu)_c > 1$ for small $\gamma$, decreases
with increasing $\gamma$, equals unity at $\gamma = \gamma_P
= \zeta_1^{-1}(0) - 1 \approx 0.69$, changes sign at 
$\gamma = \gamma_* \approx 3.9$, and 
is negative for $\gamma > \gamma_*$.  
(One should not be alarmed by this sign change:  the
concentration of up particles in the normal state is larger
than that of the down particles if
$m_{\uparrow} (\mu + h) > m_{\downarrow} (\mu - h)$,
{\it i.e.}, $\frac{h}{\mu} > \frac{ 1 - \gamma}{1 + \gamma}$.)
This behavior is the result of unequal masses of the two species.
%with the chemical potential  of the majority species decreases
%with $m_{\uparrow}$.
This is particularly clear if we examine the behavior of 
$(h/\mu)_c$ for large $\gamma$.  In this limit,
the solution to eq (\ref{eqm}) is given asymptotically by
$(h/\mu)_c = -1 + c_1 / \gamma^{3/5}$ where
$c_1 = 2 \left[ (\zeta_1(0))^{-3/2} - 1 \right]^{2/5}$.
The dependence on $\gamma$ is due to the different
density of states between the two species.
With $\zeta_1(0) \approx 0.59$, we find $c_1 \approx 2.16$.
%Therefore for large mass ratio, 
%$(h/\mu)_c$ decreases towards $-1$.
  This 
asymptotic is plotted also in Fig \ref{fig:Dh}, which
gives a reasonable description to the actual solution.

\begin{figure}[tbh]
\begin{center}
\includegraphics[width=3in]{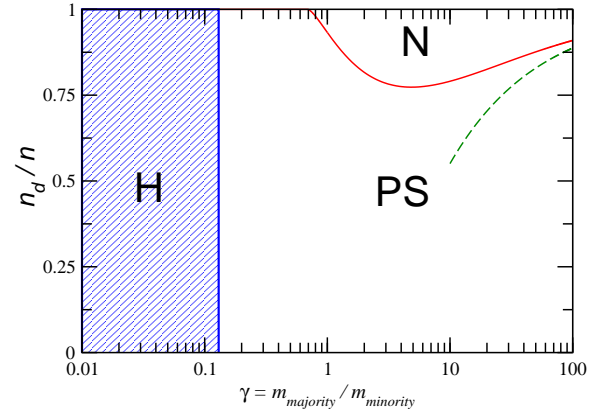}
\end{center}
%\vspace{-5pt}
 \caption{(color online) The bulk phase diagram at resonance. 
  For $n_d/n = 0$, the system is a superfluid, whereas
  it is normal for $n_d/n = 1$.  For $0 < n_d/n < 1$,
 the phases are: H - the homogeneous phase,
  PS - phase separated state, N - normal phase.
 The dashed line is the asymptotic formula discussed in text.}
 \label{fig:nd}
 \vspace{-5pt}
\end{figure}

The ratio of the minority to majority density is
$(n_{\downarrow}/n_{\uparrow})_N =$
${\cal H} \left[ ( 1 - (h/\mu)_c )/ ( \gamma ( 1 + (h/\mu)_c) ) \right]^{3/2}$.
%Defining $n_d = n_{\uparrow} - n_{\downarrow}$,
$(n_d/n)_N$ can be obtained via $ ( 1 - (n_{\downarrow}/n_{\uparrow})_N)/ 
(1 + (n_{\downarrow}/n_{\uparrow})_N) $.
This ratio as a function of $\gamma$ is plotted in
Fig \ref{fig:nd}.
$(n_d/n)_N = 1$ for $\gamma < \gamma_P$.  There
the only stable normal phase is the completely polarized phase.
  Since 
$n_{\downarrow} =n_{\uparrow}$ in the superfluid phase,
$(n_d/n)_N$ is also the maximum allowed $(n_d/n)$ for
the phase separated system.  For 
$ 0 < n_d/n < (n_d/n)_N$, the system phase separates
into a superfluid and a normal state, whereas
for $ (n_d/n) > (n_d/n)_N$ the system is in the normal state.
These two different regions are labelled as PS and N in Fig \ref{fig:nd}.
The dependence of $(n_d/n)_N$ on $\gamma$ is a result
of competition between the $\gamma$ dependence of $(h/\mu)_c$
and that of the density of states.  For large $\gamma$, using
the asymptotic form for $(h/\mu)_c$ discussed before,
we get $(n_{\downarrow}/n_{\uparrow})_N \approx
(2/c_1)^{3/2} / \gamma^{3/5}$, and thus
$(n_d/n)_N \approx 1 -
  2^{5/2} / (c_1^{3/2} \gamma^{3/5}) $.
This asymptotic dependence is also plotted in Fig \ref{fig:nd}.
The dependence of
$(n_{\downarrow}/n_{\uparrow})_N$ on $\gamma$ is a combined
result of $ ( 1 + (h/\mu)_c ) \propto 1 / \gamma^{3/5}$ and
the ratio of density of states of the down to up particles
($\propto \gamma^{-1}$).
This competition also results in the non-monotonic behavior
of $(n_d/n)_N$ on $\gamma$, with a minimum occuring at
$\gamma$ slightly larger than $\gamma_*$.

The density of the normal fluid is given by
$n_N = \frac{1}{6 \pi^2}
\left[ ( 2 m_{\uparrow} ( \mu + h))^{3/2} +
 (2 m_{\downarrow} ( \mu - h))^{3/2}  {\cal H} \right]$
and thus its ratio to that of the superfluid state is
\begin{equation}
\left( \frac{n_N}{n_S} \right) =
\frac{\zeta_1^{3/2}(0)}{2}
\left[ \left( \frac{1 + \gamma}{2} \right)^{3/2}
    \left( 1 + \left( \frac{h}{\mu} \right)_c \right)^{3/2} + 
   \left( \frac{1 + \gamma^{-1}}{2} \right)^{3/2}
    \left( 1 - \left( \frac{h}{\mu} \right)_c \right)^{3/2} {\cal H}
   \right]
\label{nratio}
\end{equation}

\begin{figure}[tbh]
\begin{center}
\includegraphics[width=3in]{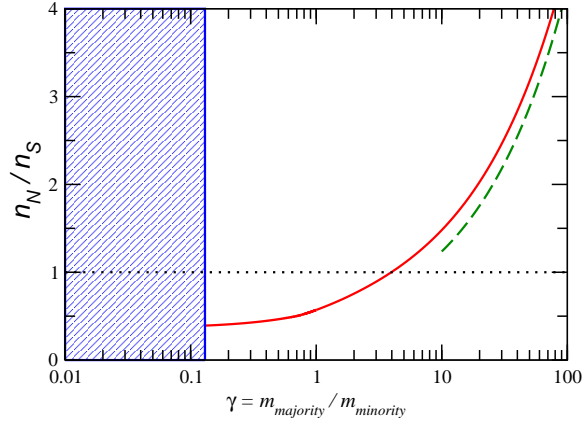}
\end{center}
%\vspace{-5pt}
 \caption{(color online) The ratio of $n_N$ to $n_S$ at phase equilibrium
  for the phase separated region.
The dashed line is the asymptotic formula discussed in text.}
 \label{fig:nratio}
 \vspace{-5pt}
\end{figure}

The dependence of 
$n_N/n_S$ as a function of $\gamma$ is
as shown in Fig \ref{fig:nratio}.  
$n_N/n_S$ increases with increasing $\gamma$.
For $\gamma < \gamma_* \approx 3.9$, $n_N/n_S < 1$,
but for $\gamma > \gamma_*$, $n_N/n_S > 1$.
At $\gamma_*$ where $(h/\mu)_c = 0$, 
$n_N/n_S$ is exactly unity as can be verified easily
by comparing eqs (\ref{eqm}) and (\ref{nratio}).
For large $\gamma$, we have
$n_N/n_S \sim \zeta_1^{3/2}(0) [ c_1^{3/2} \gamma^{3/5} + 2^{3/2} ] /
          2^{5/2}$, hence 
$n_N/n_S$ increases with $\gamma$ roughly as
$\gamma^{3/5}$.  This dependence is again a combination
of $\gamma$ dependence of $(h/\mu)_c$ and the density of
states factors.  The asymptotic behavior is also plotted
in Fig \ref{fig:nratio}.

\begin{figure}[tbh]
\begin{center}
\includegraphics[width=3in]{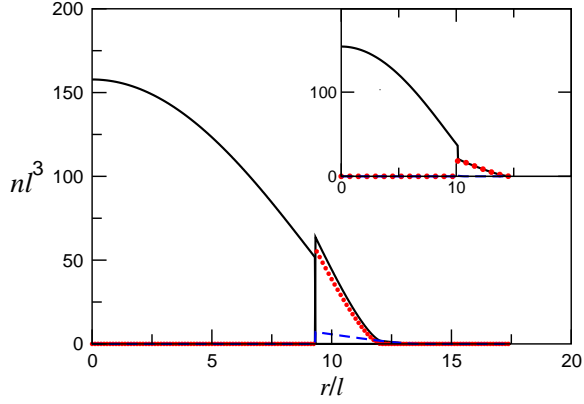}
\end{center}
%\vspace{-5pt}
%\special{psfile=den.eps hoffset = 20 hscale=42 vscale=42 angle=-90}
%\vspace*{3.5in}
 \caption{(color online) Density profile at resonance for
$N= 4.1 \times 10^{5}$, $N_d = 6.6 \times 10^{4}$, $N_d/N = 0.16$.
$\alpha_{\uparrow} = \alpha_{\downarrow} = \alpha$, 
$\gamma = 6.7$. 
Full line:  total density,
 dots and dashed:  $n_{\uparrow,(\downarrow)}$ 
in the normal fluid region. Here $r$ is normalized to $l \equiv
1/[ \alpha (2 m_r)]^{1/4}$ and the density $n$ to $l^{-3}$.
Inset: density profile for the same $N$ and $N_d$ for $\gamma=1$.
$n_{\downarrow}(r)$ for the normal phase is very small and thus
only barely visible.
The density profiles with the same $N_d/N$ but $N_{\rm new} = \beta N$
can be obtained by the scaling relation
$n_{\rm new} (\beta^{1/6} r) = \beta^{1/2} n(r)$.}
 \label{fig:den}
 \vspace{-5pt}
\end{figure}

The ratio $n_N/n_S$ and $(n_d/n)_N$ above are directly
reflected in the density profile of a trapped gas.
Let us consider for simplicity an isotropic harmonic
oscillator trap, with trap potential energy for the two
species $ \frac{1}{2} \alpha_{\sigma} r^2$.
The local chemical potentials are given by
$\mu_{\sigma} (r) = \mu_{\sigma}^o - \frac{1}{2} \alpha_{\sigma} r^2$,
and hence
$\mu(r) = \mu^o - \frac{1}{4} (\alpha_{\uparrow} + \alpha_{\downarrow}) r^2$
and
$h(r) = h^o - \frac{1}{4} (\alpha_{\uparrow} - \alpha_{\downarrow}) r^2$.
Generally $h(r)/\mu(r)$ increases from the center towards outside.
Therefore a superfluid core exists for $r < r_c$
where $r_c$ is defined by
$h(r_c)/\mu(r_c) = (h/\mu)_c$.  In this region,
the local density $n(r)$ is given by
$( 4 m_r \mu(r) )^{3/2}/ ( 3 \pi^2 \zeta_1^{3/2}(0))$
with $n_{\uparrow}(r) = n_{\downarrow}(r) = n(r)/2$.
For $r > r_c$ we have a normal fluid.  There
$n_{\sigma}(r) = [ 2 m_{\sigma} (\mu (r) \pm h (r)) ]^{3/2}/ ( 6 \pi^2)$
provided $ \mu(r) \pm h(r)  > 0$ and vanishes otherwise.
As an example, we show in Fig \ref{fig:den} the density
profile for the case $\alpha_{\uparrow} = \alpha_{\downarrow} =\alpha$,
$\gamma = 6.7 $ ($\approx m(^{40}{\rm K})/ m(^6{\rm Li})$).
This figure clearly shows that
the density at $r = r_c^+$ is larger than that of $r_c^-$
(with ratio $\approx 1.2$, {\it c.f.} Fig \ref{fig:nratio};
also $(n_{\downarrow}/n_{\uparrow})|_{r_c^+} \approx 0.12$
corresponding to $(n_d/n)_N \approx 0.78$, {\it c.f.}
Fig \ref{fig:nd}).
For comparison, the density profile for the case of equal mass 
($\gamma  = 1$) is shown in the inset of Fig \ref{fig:den},
which shows a density {\em drop} from the superfluid
core to the normal region.
We also note that in our present case, the minority
particles have a larger cloud radius than the majority.
This is because of the lighter mass of the minority species,
so that in fact $\mu^o_{\downarrow} > \mu^o_{\uparrow}$ 
and $h^o < 0$.

In conclusion, we have studied the superfluid pairing between
unequal mass Fermions, in particular at resonance.  We show
that, for equal populations, the thermodynamic properties of the system
are the same as the equal mass case except for
a simple rescaling, whereas for unequal populations,
they depend crucially on the mass ratio of the two species.
There is an abundance of stable or long-lived
isotopes with very different masses even within the alkalis,
ranging from $^2{\rm H}$ to $^{134}{\rm Cs}$,
and thus
ample opportunities to test the present predications.
For example, at resonance, a $^2{\rm H}$(majority)-
$^{40}{\rm K}$(minority) mixture would not phase separate,
whereas a
$^{40}{\rm K}$(majority)- $^{2}{\rm H}$(minority) mixture would.
For  $^{6}{\rm Li}$(majority)-$^{40}{\rm K}$(minority)
mixture, the system would phase separate with a density
drop from the superfluid core to a (completely polarized) normal region,
(in contrast to equal mass case, where the normal region
is only partially polarized),
whereas $^{40}{\rm K}$(majority)-$^{6}{\rm Li}$(minority)
mixture would have a density jump.

Very recently, another preprint \cite{Iskin} appears
which also deals with unequal mass Fermions.  
This research was supported by the NSC of
Taiwan under grant numbers 
NSC94-2112-M-194-008 (STW), 
NSC94-2112-M-194-001 (CHP),
and NSC94-2112-M-001-002 (SKY).
%, with additional support from National Center for Theoretical Sciences,
%Hsinchu, Taiwan.

%%%%%%%%%%%%%%%%%%%%%%%%%%%%%%%%%%%%%%%%%%%%%%%%%%%%%%%%%%%%%%%%

\newpage

%%%%%%%%%%%%%%%%%%%%%%%%%%%%%%%%%%%%%%%%%%%%%%%%%%%%%%%%%%%%%%%%%%%%%%%%%%%%%%%%%%%%%%
\vspace{15pt}

%%%%%%%%%%%%%%%%%%%%%%%%%%%%%%%%%%%%%%%%%%%%%%%%%%%%%%%%%%%%%%%%%%%%%%%%%%%%%%%%%%%%%%%
\vskip 1 cm
%%%%%%%%%%%%%%%%%%%%%%%%%%%%%%%%%%%%%%%%%%%%%%%%%%%%%%%%%%%%%%%%%%%%%%%%%%%%%%%%%%%%%%
\vspace{20pt}

%%%%%%%%%%%%%%%%%%%%%%%%%%%%%%%%%%%%%%%%%%%%%%%%%%%%%%%%%%%%%%%%%%%%%%%%%%%%%%%%%%%%%%%

%%%%%%%%%%%%%%%%%%%%%%%%%%%%%%%%%%%%%%%%%%%%%%%%%%%%%%%%%%%%%%%%%%%%%%%%%%%%%%%%%%%%%%
\vspace{15pt}

%%%%%%%%%%%%%%%%%%%%%%%%%%%%%%%%%%%%%%%%%%%%%%%%%%%%%%%%%%%%%%%%%%%%%%%%%%%%%%%%%%%%%%%
\newpage

%%%%%%%%%%%%%%%%%%%%%%%%%%%%%%%%%%%%%%%%%%%%%%%%%%%%%%%%%%%%%%%%%%%%%%%%%%%%%%%%%%%%%%
\vspace{18pt}

%%%%%%%%%%%%%%%%%%%%%%%%%%%%%%%%%%%%%%%%%%%%%%%%%%%%%%%%%%%%%%%%%%%%%%%%%%%%%%%%%%%%%%%


\begin{thebibliography}{99}
\bibitem{free} Explicitly, this free energy is 
$\Omega(\Delta,\mu,h)
 = \sum_{\vec k} [ \frac{\hbar^2 k^2} { 4 m_r} - \mu - \tilde E(k) 
 + \frac{2 m_r}{\hbar^2 k^2} |\Delta|^2 + 
   E_{\uparrow} (k) f (E_{\uparrow}(k)) ]
 - \frac{m_r}{2 \pi \hbar^2 a} |\Delta|^2$ where $f$ is
the Fermi function, and $E_{\sigma}(k)$, $\tilde E(k)$, $\mu$ are
defined in text.
\end{thebibliography}
\end{document}